\documentclass[fleqn,10pt,twocolumn]{wlscirep}
\pdfoutput=1
\title{Spontaneous Parametric Down-Conversion Induced by Non-Degenerate Three-Wave Mixing in a Scanning MEMS Micro Mirror}

\author[1,*]{Ulrike Nabholz}
\author[1]{Frank Schatz}
\author[2]{Jan E. Mehner}
\author[1,+]{Peter Degenfeld-Schonburg}
\affil[1]{Robert Bosch GmbH, Corporate Research, 71272 Renningen, Germany}
\affil[2]{Chemnitz University of Technology, 09107 Chemnitz, Germany}

\affil[*]{ulrike.nabholz@de.bosch.com}
\affil[+]{peter.degenfeld-schonburg@de.bosch.com}

\usepackage{xcolor}
\usepackage{IEEEtrantools}
\usepackage{multirow}
\usepackage{cite}

\begin{abstract}
Scanning micro-mirror actuators are silicon-based oscillatory micro-electro-mechanical systems (MEMS). They enable laser distance measurements for automotive LIDAR applications as well as projection modules for the consumer market. For MEMS applications, the geometric structure is typically designed to serve a number of functional requirements. Most importantly, the mode spectrum contains a single high-Q mode, the drive mode, which per design is expected to yield the only resonantly excited geometric motion during operation. Yet here, we report on the observation of a resonant three-mode excitation via a process known as spontaneous parametric down-conversion. We show that this phenomenon, most extensively studied in the field of nonlinear optics, originates from three-wave coupling induced by geometric nonlinearities. In combination with further Duffing-type nonlinearities, the micro mirror displays a variety of nonlinear dynamical behaviour ranging from stationary state bifurcations to dynamical instabilities observable via amplitude modulations. We are able to explain and emulate all experimental observations using a single fundamental model. In particular, our analysis allows us to understand the conditions for the onset of three-wave down-conversion which if not accounted for in the design of the MEMS structure, can have drastic impact on its functionality even leading to fracture. 
\end{abstract}
\begin{document}

\flushbottom
\maketitle
%
%
\thispagestyle{empty}

\section*{Introduction}
Micro-electro-mechanical systems (MEMS) are small-scale devices containing mechanical and electrical components \cite{Senturia2007, Korvink2006, Allen2005, Muller1991}. They are fabricated using integrated circuit and micromachining technologies \cite{Madou2002, Menz2008, Wu1997}. A scanning micro mirror denotes an example of a silicon-based MEMS actuator, often driven at resonance, where one mode of oscillation causes a reflective structure to oscillate and deflect a laser beam \cite{Li2018, Solgaard2014, Petersen1980, Kurth1998, Holmstroem2014, Schenk2014, Ye2017, Baran2012a}. For this application, requirements pertaining to the deflection angle that can be reached are of great importance.\\
In this work, we report on experimental observations showing a drastic difference between the attainable oscillation amplitudes of scanning micro mirror devices despite their matching design layout. While some micro mirrors reach the necessary deflection angles as expected, others show unstable oscillation amplitudes and can even fracture far below the deflection angles needed for reliable operation. We are able to argue that small statistically distributed differences in the mode spectrum of the mirrors originating from the process tolerances of surface micromachining, lead to drastically different device behaviour. We explain all our observations exploiting a fundamental nonlinear dynamics model known as spontaneous parametric down-conversion (SPDC).\\
SPDC has been widely studied in the context of nonlinear optics \cite{Boyd2008}. It has triggered both technological advances \cite{Ligo2011}, such as tunable lasers or high-quality squeezed light sources \cite{Takeno2007, Vahlbruch2008}, and fundamental insights in the field of quantum optics \cite{Walls2008, Carmichael2009, Degenfeld-Schonburg2015} ranging from quantum information \cite{Braunstein2005, Weedbrook2012, Leghtas2015, Foertsch2013} to modern hybrid opto-mechanical devices \cite{Aspelmeyer2014, Degenfeld-Schonburg2016}. Remarkably, we find that the underlying concepts of SPDC apply in full analogy to a complex mechanical system with practical application such as our micro mirror design.\\
In the mechanical domain, in contrast to the photonic case, the inherent geometric nonlinearities of the structure are responsible for the nonlinear coupling of vibrational modes \cite{ Nayfeh2008, Strogatz2007}. 
Nonlinear dynamics of mechanically coupled oscillators have long been an important field of study with a broad range of engineering applications. As MEMS devices grew smaller in size and more complex in design, their nonlinear dynamic behavior sparked interest and thus the development of MEMS-specific models emerged \cite{ Lifshitz2009, Younis2003, Mestrom2008, Kacem2009, Najar2010} and is still ongoing, discussing complex nonlinear dynamic phenomena, such as limit cycles \cite{ Zaitsev2012, Aubin2004}, under various internal resonance conditions. 
%
\subsection*{Micro mirror design and actuation principle}
In general, a micro mirror can be described as a scanning system with a torsional degree of freedom utilizing mechanical structures \cite{Motamedi2005}.\\
\begin{figure}[ht]
	\centering
	\includegraphics[width=1\linewidth]{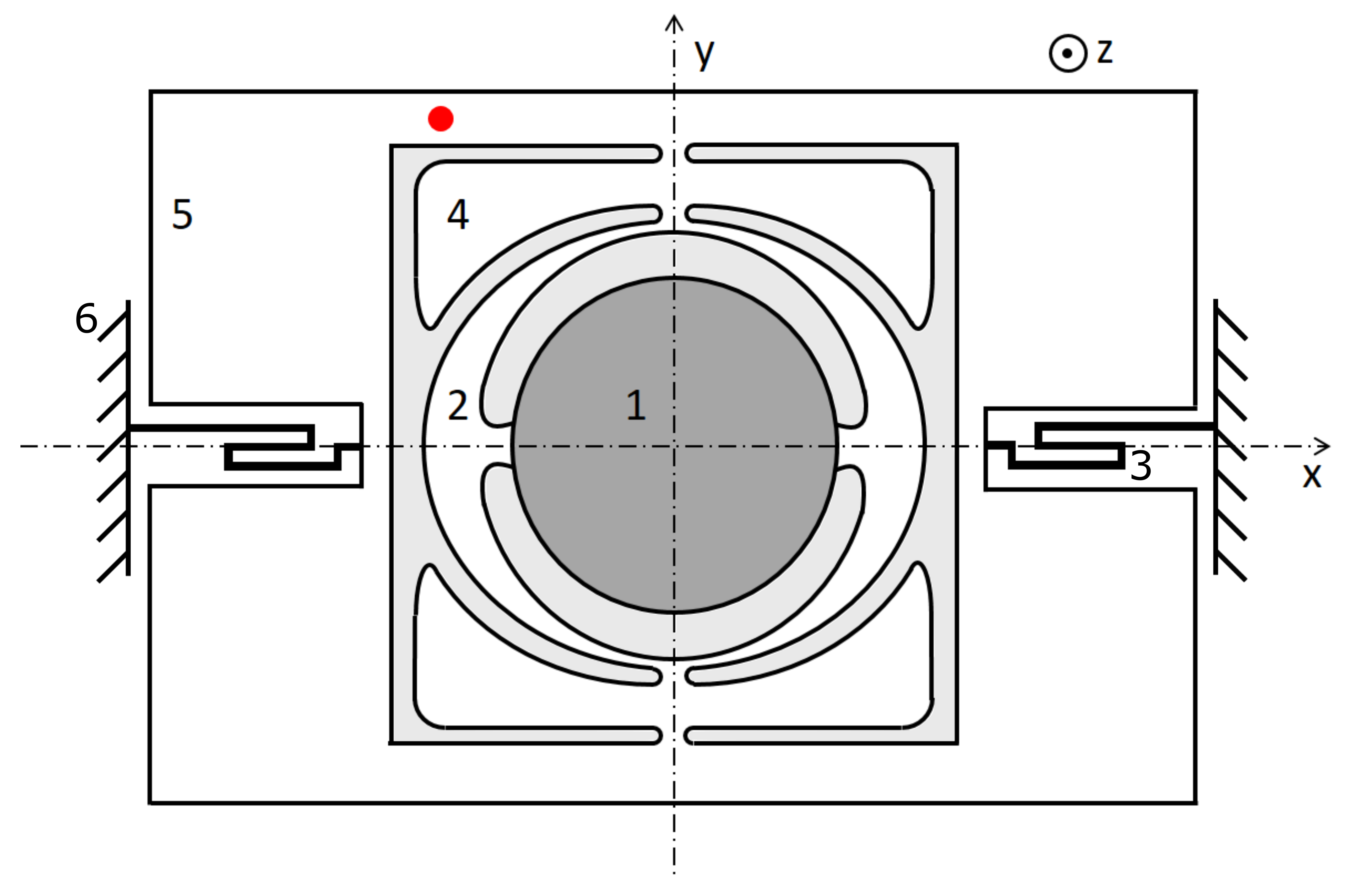}
	\caption{Schematic drawing of the planar micro mirror design layout. 1: reflective structure, 2: springs of the resonant axis, 3: springs of the static axis, 4: masses, 5: frame. A coordinate system used to describe the mode shapes is indicated. The red dot denotes the point of measurement used for Fig.\,\ref{fig:freq_spectrum_ldv}.}
	\label{fig:mirrorsketch}
\end{figure}
\noindent
Fig.\,\ref{fig:mirrorsketch} shows a schematic drawing of our micro mirror illustrating its working principle. For scanning applications, the mirror has two main axes of deflection or rather rotation: the static axis in x-direction around which the mirror typically performs the slow (off-resonant) rotation and the resonant axis in y-direction around which the mirror performs the fast rotation. In this paper we focus on the resonantly actuated drive mode of the  fast oscillation around the resonant axis. Thus, in all measurements, the slow axis is not actuated at all.\\
In the drive mode, the reflective structure (1) and the surrounding springs (2) oscillate or rather rotate in phase around the resonant axis in order to deflect a laser beam by a defined angle. The rotation occurs out-of-phase with the out-of-plane motion of the masses (4). The parasitic modes identified from the measurements in Fig.\,\ref{fig:freq_spectrum_ldv} have distinctly different mode shapes: Parasitic mode 1 is a so-called in-plane mode since the main movement of regions 1, 2, 4 and 5 occurs in phase and in the x-y-plane. It mainly causes deformations of the springs of the static axis (3) which are connected to the frame (5) and the anchor (6). The amount of deflection in z-direction is small, but its motion is similar to the rotation of the drive mode. Parasitic mode 2 is an out-of-plane mode where the reflective structure (1) and the resonant springs (2) move along the z-axis out of phase with the four distributed masses (4). \\
The drive mode actuation is implemented using a piezoelectric actuation principle \cite{Baran2012a}, thus, the actuation force is expected to be linear over deflection amplitude and input voltage in contrast to a typical electrostatic actuation principle \cite{Tang1990}. Indeed, we have confirmed that the critical deflection angles of SPDC do not depend on the applied voltage settings. Moreover, we can also expect material nonlinearities to be negligible since silicon shows a linear-elastic stress-strain dependency for strains within the reach of our experiments and design goals \cite{Petersen1982}.
\subsection*{The phenomenology of SPDC}
\begin{figure}[thbp]
	\centering
	\includegraphics[width=1.0\linewidth]{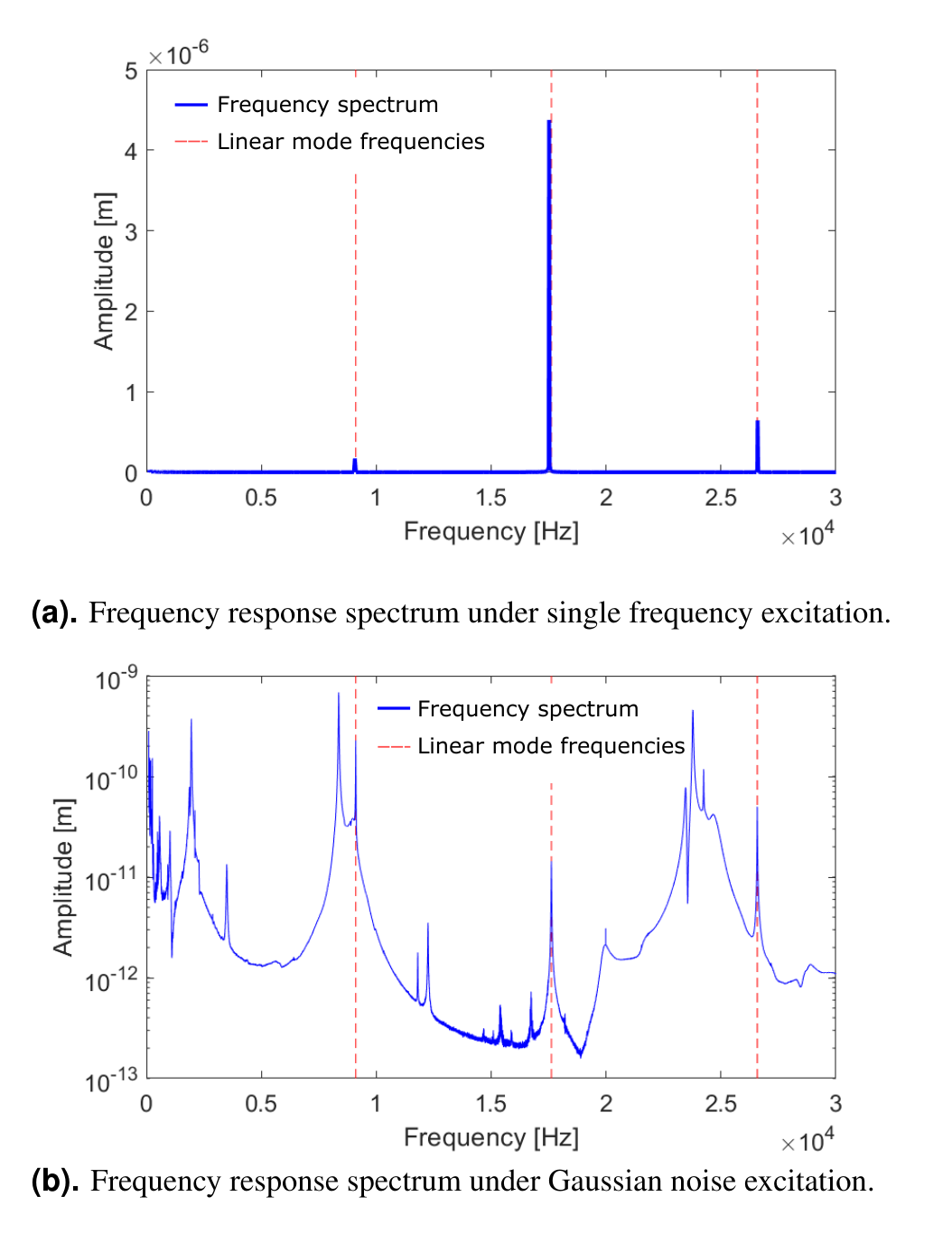}
	\caption{Laser Doppler vibrometry measurements of micro mirror device A. In both cases, the dashed red lines mark the linear mode frequencies of the three relevant modes: At the lowest \(f_{0,1} = 9126\,Hz\) and middle \(f_{0,2} = 17673\,Hz\) frequencies lie the parasitic modes labelled 1 and 2, at the highest frequency \(f_{0,0} = 26712\,Hz\) the drive mode. In Fig.\,\ref{fig:freq_spectrum_ldv}a, the system is actuated sinusoidally at a drive frequency \(f_{\mathrm{d}}\), in this case \(f_{\mathrm{d}} = 26712\,Hz\) leading to pronounced resonance peaks. Apart from the drive mode, the two parasitic modes resonate at frequencies that add up exactly to the value of the drive frequency but only above the critical deflection angle needed for SPDC. The measurement is performed with the laser spot on the frame of the micro mirror (denoted by the red dot shown in Fig.\,\ref{fig:mirrorsketch}). Thus, the oscillation amplitude is highest for parasitic mode 2 that exhibits large deformations of the frame structure. Fig.\,\ref{fig:freq_spectrum_ldv}b shows the amplitude response curve that results from vibratory Gaussian noise actuation in out-of-plane direction. The prominence and shape of the peaks varies due to the superposition of modes in the complex MEMS structure which influences the quality factors of the modes and the ease of actuation in out-of-plane direction. To better depict modes that only reach small amplitudes, the amplitude is plotted logarithmically.}
	\label{fig:freq_spectrum_ldv}	
\end{figure}

In the case presented here, the mode spectrum of the micro mirror design contains a 1:1:1 internal resonance, meaning that the sum of the resonance frequencies of two modes \(f_{0,1}\) and \(f_{0,2}\) approximately adds up to the resonance frequency \(f_{0,0}\) of the drive mode, i.e. \(f_{0,0} \approx f_{0,1} + f_{0,2}\). Fig.\,\ref{fig:freq_spectrum_ldv}b shows a typical mode spectrum displayed by the response spectrum of a micro mirror device under Gaussian noise excitation. The measurement was performed using Laser Doppler vibrometry (LDV) for detection and a vibratory plate below the device for mechanical actuation. \\
\noindent
Statistical distributions of resonance frequencies due to process tolerances of surface micromachining govern the quantity \(f_{0,0} - \left(f_{0,1} + f_{0,2} \right)\) which, as we will show in the course of this paper, mainly influences the critical deflection angles needed for the onset of parametric down-conversion. As a consequence, some devices display SPDC already for small deflection angles, whereas others behave as expected within the relevant parameter range. \\
\begin{figure}[htbp]
	\centering
	
	\includegraphics[width=1.0\linewidth]{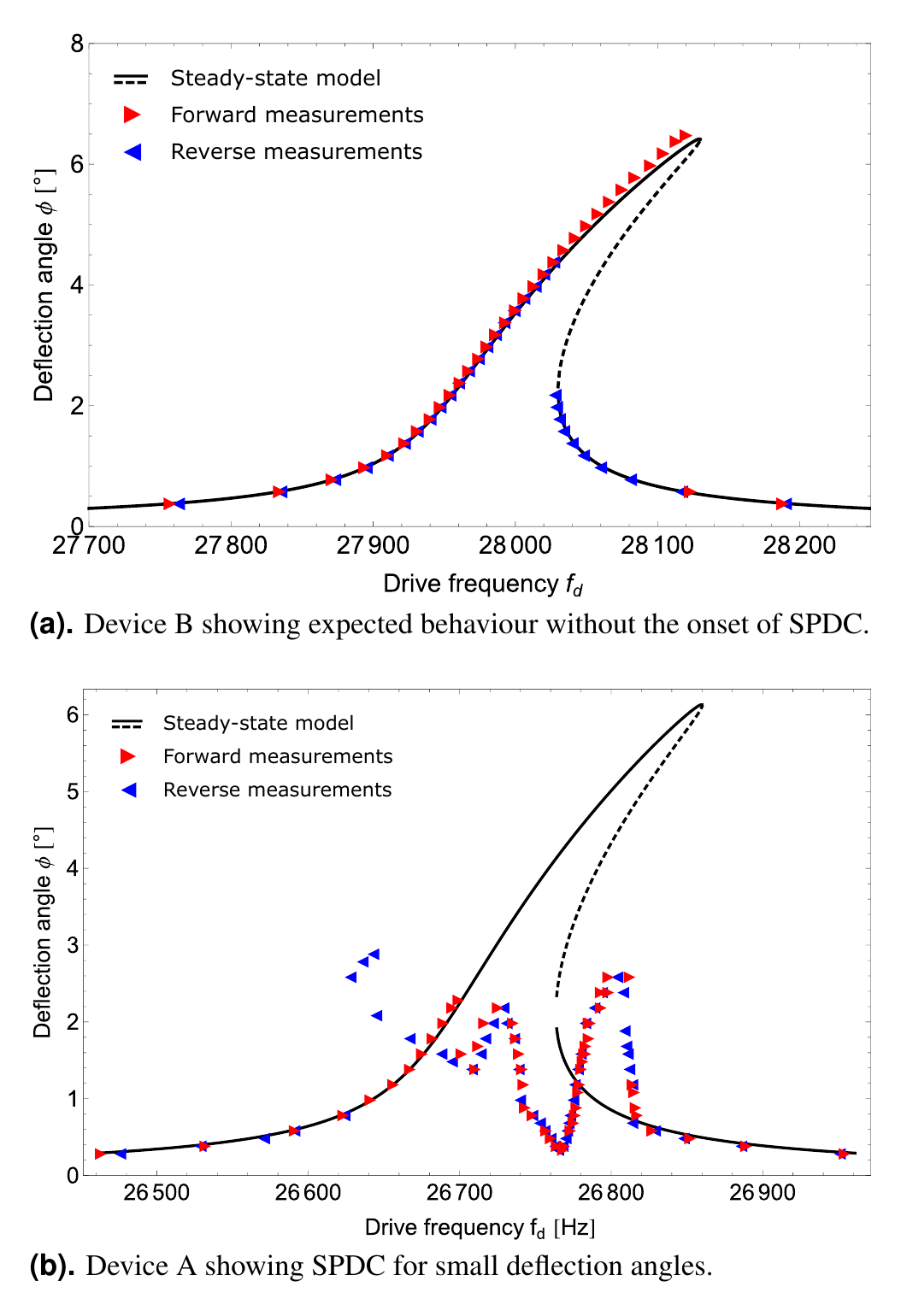}
	\caption{Optical measurements of amplitude response curves. Triangles denote the measured data, the forward frequency sweep is depicted in red, the reverse sweep in blue.  The steady-state model for a nonlinearly damped Duffing oscillator \cite{Nabholz2018} is shown in black. In Fig.\,\ref{fig:comparison_mode_coupling}a, the oscillation can be accurately modelled using a single mode of oscillation \cite{Nabholz2018}, whereas in Fig.\,\ref{fig:comparison_mode_coupling}b, the mirror displays SPDC where the single mode description does not suffice.}
	\label{fig:comparison_mode_coupling}	
\end{figure}
\noindent
Fig.\,\ref{fig:comparison_mode_coupling}a displays such an expected behaviour in the form of a Duffing-like \cite{Duffing1918} amplitude-response curve measured for the micro mirror device B with \(f_{0,0} - \left(f_{0,1}+f_{0,2}\right)= -150\,Hz\). In stark contrast, the measurement shown in Fig.\,\ref{fig:comparison_mode_coupling}b where the resonance condition of the micro mirror device A is found to be \(f_{0,0}-\left(f_{0,1}+f_{0,2}\right)=-87\,Hz\) differs greatly from the expectation. Besides the pronounced energy depletion of the drive mode, above the so called critical point or rather threshold of SPDC, we also find its characteristic feature in the response spectrum of the micro mirror. The measured spectral response under resonant actuation of the mirror by a sinusoidal force with frequency \(f_{\mathrm{d}}\) is shown in Fig.\,\ref{fig:freq_spectrum_ldv}a. It exhibits sharp peaks not only at the drive frequency \(f_{\mathrm{d}}\) but also at frequencies \(f_{osc,1} \approx f_{0,1}\) and \(f_{osc,2} \approx f_{0,2}\) such that \(f_{\mathrm{d}} - \left(f_{osc,1}+f_{osc,2} \right) = 0\).\\
Throughout this paper we will present a detailed analysis of the model for SPDC where the micro mirror will be modelled as an externally driven and damped Duffing oscillator \cite{Duffing1918} coupled with two parasitic modes through a three-wave coupling term. Additionally, since the mirror is actuated at ambient pressure, a nonlinear damping term is included for all three mechanical modes \cite{Nabholz2018}. With such a simple model, we are able to simulate and thus explain the observed effects including bifurcations, resonant actuation of parasitic modes, amplitude depletion of the drive mode, hysteretic behaviour, critical slowing-down \cite{Degenfeld-Schonburg2015, Degenfeld-Schonburg2016} and even limit cycles, i.e. amplitude modulations in the stationary state \cite{Strogatz2007}. Most importantly, we provide an analytic expression for the critical deflection angles. For engineering applications, the product specifications include a required deflection angle. The critical deflection angle allows us to differentiate between devices that fulfil the specification and devices that do not. Thus, it is of highest practical importance. 
\section*{Results}
\subsection*{Modal three-wave coupling induced by geometric nonlinearities}
\label{spdc}
We will now briefly discuss some fundamental properties of elasticity theory to argue for the existence of the essential physical process underlying SPDC given by a three-mode or rather three-wave interaction. Such an interaction is described by the nonlinear three-wave energy term
\begin{equation}
U_{\mathrm{SPDC}}=\tilde{\alpha} q_0 q_1 q_2, 
\end{equation}
where \(\tilde{\alpha}\) denotes the coupling strength of the three-wave coupling between the vibrational modal amplitudes \(q_0, q_1, q_2\) of the mechanical structure. As before, the index '0' denotes the drive mode and the indices '1' and '2' the first and second parasitic mode, respectively. The magnitude of the resonance frequencies is given as \(f_{0,1} < f_{0,2} < f_{0,0}\). 
In elasticity theory, the internal energy of any geometrical structure is known as the strain energy. In the total Lagrangian description of continuum mechanics \cite{Kim2014, Sathyamoorthy1997} it is given by 
\begin{equation}
U_{\mathrm{strain}}=\frac{1}{2} \int\limits_{V_0} \mathrm{Tr} \, \left\{\textbf{S} \left(\vec{x},t\right) \textbf{E}\left(\vec{x},t\right)\right\} d^3 x, 
\end{equation}
with \(V_0\)  denoting the volume of the undeformed structure, \(\textbf{Tr}\) the trace operation and \(\textbf{E}\) and \(\textbf{S}\) the strain- and the stress-field tensor, respectively, depending on the material point \(\vec{x}\) as well as time \(t\). In particular, the strain-field \(E\) is given by the so called Green-Lagrange strain measure \cite{Kim2014, Sathyamoorthy1997} defined by its tensor matrix elements
\begin{equation}
E_{nm} = \frac{1}{2} \left(\frac{\partial u_n\left(\vec{x},t\right)}{\partial x_m} + \frac{\partial u_m\left(\vec{x},t\right)}{\partial x_n} + \sum_{l=1}^{3} \frac{\partial u_l\left(\vec{x},t\right)}{\partial x_n} \frac{\partial u_l\left(\vec{x},t\right)}{\partial x_m}\right)
\end{equation}
with the displacement vector field \(\vec{u}\left(\vec{x},t\right)\) at every material point \(\vec{x}\) and time \(t\). The strain term, quadratic in the displacement, accounts for large structural deformations and describes what is known as \textit{geometric nonlinearity}. Typically, for small deformations this term is neglected \cite{Landau1991}. Furthermore, the stress field is given by the so called second Piola-Kirchhoff stress which in the linear elastic regime obeys the constitutive relation \(S_{nm} = D_{nmlk} E_{lk}\)  with the constant fourth-order material tensor \(D_{nmlk}\)  depending on Young's Moduli and Poisson's ratios of the material.\\
In summary, it is important to understand that the strain energy \(U_{\mathrm{strain}}\) including geometric nonlinearities contains cubic and quartic terms in the displacement vector \(\vec{u}\left(\vec{x},t\right)\) even in the linear elastic material regime. Finally, when we expand the displacement vector field in a suitable basis \(\vec{u}\left(\vec{x},t\right)=\sum_{n=1} f_n \left(\vec{x}\right) q_n \left(t\right)\), such as the basis formed by the vibrational modes \(f_n \left(\vec{x}\right)\) of the structure, we find that an energy term of the form \(U_{\mathrm{SPDC}}\) is indeed present in the strain energy describing geometrically nonlinear structural behaviour. Thus, we conclude that the SPDC effects observed in our micro mirror design originate from geometric nonlinearities.\\ %
\subsection*{System model} 
%
%
%
%
In full generality, the modal representation of the strain energy is given by
\begin{equation}
U_{\mathrm{strain}}\!=\!\sum_{n}\!\frac{1}{2} \omega_{0,n}^2 q_n^2\!+\!\sum_{n,m,l}\!\tilde{\alpha}_{n,m,l}q_n q_m q_l \!+\!\sum_{n,m,l,k}\!\tilde{\beta}_{n,m,l,k}\!q_n q_m q_l q_k \label{eq:Ustrain}
\end{equation}
and includes three- and four-wave terms of the form \(\alpha_{n,m,l} q_n q_m q_l\) and \(\beta_{n,m,l,k} q_n q_m q_l q_k\), respectively \cite{Touze2014}.\\
Out of all possible three- and four-wave mixings, we consider only the resonant terms between the drive mode and the two parasitic modes since these are the most relevant ones in our high-Q system. This procedure is based on the fact that nonlinear terms depend on the parasitic mode amplitudes which are small and therefore negligible unless the internal resonance conditions enable a resonant excitation. In such cases, the amplitudes of the parasitic modes can be enlarged by the high quality factors and therefore the nonlinear terms can no longer be neglected.\\
Within the scope of this paper, we focus on the experimental (instead of the numerical) parameter extraction for the relevant couplings \(\alpha_{n,m,l}\) and \(\beta_{n,m,l,k}\) using a three degree-of-freedom model. We describe the details of our parameter extraction procedure in a later section.\\
Most importantly, we consider energy terms of the form \(\alpha_{n,m,l} q_n q_m q_l\) with \(n \neq m \neq l\), which are relevant whenever a 1:1:1 internal resonance is present, that is whenever \(f_n \approx f_m +f_l\) is fulfilled. (Only then, the averaging procedure which we will introduce later and describe in detail in the Methods section yields an influence of these terms. Otherwise, the terms can be neglected.)
Furthermore, we consider terms that are always resonant regardless of the mode spectrum, namely the Duffing nonlinearity of a mode \(n\) given by \(\beta_{n,n,n,n} q_n^4\) and terms of the form \(\beta_{n,n,m,m} q_n^2 q_m^2\)  with \(n \neq m\) to account for frequency shifts of the n-th mode resonance frequency as a function of the m-th mode oscillation amplitude \cite{Lulla2012}.

%
\noindent
Note that all possible permutations of the indices have to be included. For the sake of clarity, the notations are simplified and are given by the following relations: The Duffing nonlinearity is written as \(\tilde{\beta_n} := 4 \beta_{n,n,n,n}\), the three-wave coupling \(\tilde{\alpha} := \alpha_{n,m,l} + \alpha_{n,l,m} + \alpha_{m,n,l} + \alpha_{m,l,n} + \alpha_{l,n,m} + \alpha_{l,m,n}\) and the mutual frequency shifts between modes which we will call cross-Duffing nonlinearities as \(V_{n,m} := 4\beta_{n,n,m,m} + 8\beta_{n,m,n,m}\). The cross-Duffing terms, however, can be neglected since an LDV measurement, see methods section, revealed that \(V_{0,1},V_{0,2}\) are negligibly small. From this we deduce that \(V_{1,2}\) is also negligible. Due to the fact that the actuation principle of the micro mirror is not designed for a direct actuation of the parasitic modes, their actuation induces high local stress which entails the risk of fracture. Considering the limited number of available micro mirror samples, we thus refrain from attempts at measuring \(V_{1,2}\) directly.\\
In the modal representation, the equations of motion of our model are given by
\begin{IEEEeqnarray}{ll}
	\ddot{q}_0\!+\!\frac{\omega_{0,0}}{Q_0}\dot{q}_0\!+\!\omega_{0,0}^2 q_0\!+\!\tilde{\beta_{0}} q_0^3\!+\!\frac{\omega_{0,0}}{Q_{\mathrm{nl},0}}q_0^{2} \dot{q_0}\!&+\!\tilde{\alpha}\!q_1 q_2\!\nonumber\\
	& =\!F_0\!\sin\left(\omega_{\mathrm{d}} t\right) \label{eq:3dof_ode0} \\
	\ddot{q}_1\!+\!\frac{\omega_{0,1}}{Q_1}\dot{q}_1\!+\!\omega_{0,1}^2 q_1\!+\!\tilde{\beta_{1}} q_1^3\!+\!\frac{\omega_{0,1}}{Q_{\mathrm{nl},1}}q_1^{2} \dot{q_1}\!&+\!\tilde{\alpha} q_0 q_2\!=\!0 \label{eq:3dof_ode1}\\
	\ddot{q}_2\!+\!\frac{\omega_{0,2}}{Q_2}\dot{q}_2\!+\!\omega_{0,2}^2 q_2\!+\!\tilde{\beta_{2}} q_2^3\!+\!\frac{\omega_{0,2}}{Q_{\mathrm{nl},2}}q_2^{2} \dot{q_2}\!&+\!\tilde{\alpha} q_0 q_1\!=\!0 \label{eq:3dof_ode2}
\end{IEEEeqnarray}
Here, the modal amplitudes of the oscillation are \(q_0, \: q_1, \: q_2\) and the linear mode frequencies are given by \(\frac{\omega_{0,0}}{2 \pi}, \frac{\omega_{0,1}}{2 \pi}, \frac{\omega_{0,2}}{2 \pi}\). The equations of motion including the relevant three- and four-wave couplings are further expanded by phenomenological nonlinear damping terms that have been shown to influence the behaviour of such micro mirror devices \cite{Nabholz2018}. Thus, \(Q_0, \: Q_1, \:  Q_2\) denote the linear quality factors and \(Q_{\mathrm{nl},0}, \: Q_{\mathrm{nl},1}, \: Q_{\mathrm{nl},2}\) the nonlinear quality factors. The amplitude of the input force is given by \(F_0\) and the angular oscillation frequency of the drive mode by \(\omega_{\mathrm{d}} := \omega_{\mathrm{osc},0}\).\\ 
\noindent
A trivial solution of the system of nonlinear differential equations of motion exists for \(q_1 = q_2 = 0\). However, at the critical point or rather threshold (which denotes the onset of SPDC), this trivial solution becomes unstable in favour of a stable solution with \(q_1 \neq 0, \: q_2 \neq 0\). Such behaviour occurs at transcritical bifurcations, where a fixed point changes its stability \cite{Strogatz2007}. This threshold corresponds to a critical deflection angle that depends on the detuning of the linear mode frequencies and thus varies for different micro mirror devices. 
%
The critical angle can lie within the testing range or far beyond any achievable deflection. The effects that occur in such a nonlinear system can therefore be classified into three categories: System behaviour below threshold, at threshold and above threshold. In the following, each of these system states or rather system phases will be treated separately.\\
\subsubsection*{System model below threshold}
\label{1dof}
Below a threshold value of the drive mode amplitude, the parasitic modes are not actuated. The solution branch \(q_1 = q_2 = 0 \ \forall t\) provides a stable solution of equations \eqref{eq:3dof_ode1} and \eqref{eq:3dof_ode2}. Moreover, equation \eqref{eq:3dof_ode0} simplifies to the model of one forced and nonlinearly damped Duffing oscillator for which a known approximative steady-state solution exists \cite{Nayfeh1981}.
\subsubsection*{System model above threshold}
\label{3dof}
When the deflection angle is larger than its threshold value, the trivial solution branch with \(q_1 = q_2 = 0\) becomes unstable in favour of the above threshold solution with \(q_1 \neq 0 \neq q_2\) and resonant actuation mediated by the three-wave coupling occurs.
In order to arrive at a simplified system model and in order to even give an analytical expression for the critical deflection angle, an averaging procedure is carried out based on the rotating wave approximation (RWA), where only resonant terms are considered and their fast oscillations are neglected. The RWA has been widely used in the field of nonlinear optics and atomic physics, see references \cite{Boyd2008, Walls2008, Carmichael2009, Cohen1992, Fox2006} and references therein. For the sake of completeness, we will describe the methodological steps of the RWA which are relevant for this work in the Methods section. In essence, like in other averaging methods \cite{Strogatz2007}, the model is reduced to the envelope of the oscillation amplitude rather than its fast sinusoidal changes. An approximate model including all nonlinear terms in equations (\ref{eq:3dof_ode0}-\ref{eq:3dof_ode2}) can thus be derived using RWA. We have verified the accuracy of the RWA for some samples within the parameter ranges of this study by using full transient time-domain simulations of equations (\ref{eq:3dof_ode0}-\ref{eq:3dof_ode2}).\\
First, we introduce the complex coordinates \(a_n\) and \(a_n^*\). 
The relations between these coordinates and \(q_n\), \(p_n := \dot{q_n}\) are defined by
\begin{IEEEeqnarray}{lll}
q_n & \: = \: & \sqrt{\frac{1}{2\omega_{0,n}}}\left(a_n e^{-i\omega_{\mathrm{osc},n}t}+a_n^*e^{i\omega_{\mathrm{osc},n}t}\right) \label{eq:q_n}\\
p_n & \: = \: & -i\sqrt\frac{\omega_{0,n}}{2}\left(a_n e^{-i\omega_{\mathrm{osc},n}t}-a_n^* e^{i\omega_{\mathrm{osc},n}t}\right). \label{eq:p_n}
\end{IEEEeqnarray}
Next, we rescale the Duffing coefficient \(\beta_n := \frac{3 \tilde{\beta_{n}}}{\omega_{0,n}^2}\), the three-wave coupling \(\alpha := \frac{\tilde{\alpha}}{2\sqrt{2\omega_{0,n}\omega_{0,m}\omega_{0,l}}}\) and the input force 
\(P_n := \frac{i}{2\sqrt{2\omega_{0,n}}}F_n\left(t\right)\)
, and introduce the effective detunings \(b_n = \omega_{0,n} - \omega_{\mathrm{osc},n} + \beta_n |a_n|^2\) and the effective damping rate \(d_n = \frac{\omega_{0,n}}{2Q_n} + \frac{1}{4 Q_{\mathrm{nl},n}} a_n |a_n|^2\).\\
%
%
In the Methods section, we illustrate the derivation of the approximate model using the RWA. Note that for the RWA to be applicable, we have to specify the relation between the oscillation frequencies of the three modes as \(\omega_{\mathrm{osc},0} = \omega_{\mathrm{d}} = \omega_{\mathrm{osc},1} + \omega_{\mathrm{osc},2}\). \\
We obtain two first order differential equations for each complex coordinate pair of the three modes:
\begin{IEEEeqnarray}{lll}
	\dot{a}_0 & \: = \: & \left(-i b_0 - d_0\right) a_0 - i P_0^* - i\alpha a_1 a_2 \label{eq:a0} \\
	\dot{a}_0^* & \: = \: & \left(i b_0 - d_0\right) a_0^* + i P_0 + i\alpha a_1^* a_2^* \label{eq:a0s} \\
	\dot{a}_1 & \: = \: & \left(-i b_1 - d_1\right) a_1 -i\alpha a_0 a_2^* \label{eq:a1} \\
	\dot{a}_1^* & \: = \: & \left(i b_1 - d_1\right) a_1^* + i\alpha a_0^* a_2 \label{eq:a1s} \\
	\dot{a}_2 &\:=\:& \left(-i b_2 - d_2\right) a_2 - i \alpha a_0 a_1^* \\ \label{eq:a2}
	\dot{a}_2^* &\:=\:& \left(i b_2 - d_2\right) a_2^* + i\alpha a_0^* a_1 \label{eq:a2s} 
\end{IEEEeqnarray}
Equations (\ref{eq:a0}-\ref{eq:a2s}) do not include the fast oscillating terms and thus, the computing time needed to simulate a frequency sweep is vastly reduced compared to the full transient simulation of equations (\ref{eq:3dof_ode0}-\ref{eq:3dof_ode2}). Furthermore, in the steady-state with \(\dot{a}_n = \dot{a}^* = 0 \ \forall n\), the steady-state solution can be obtained directly by solving algebraic equations.  \\
\subsubsection*{System model at threshold}
\label{critical_point}
The critical amplitude separates the regions below and above threshold. The threshold can be understood as the bifurcation point \cite{Strogatz2007} at which the trivial solution becomes unstable. By exploiting the RWA, we are able to find an analytic expression for the threshold in terms of the critical oscillation amplitude of the drive mode. The strategy is to introduce the stability matrix \(L\left(t\right)\) by rewriting equations (\ref{eq:a1},\ref{eq:a2s}) into a matrix form:
%
\begin{equation}
\label{eq:stability}
\frac{\partial}{\partial t} 
\begin{pmatrix}
a_1\\a_2^*
\end{pmatrix}
= 
\underbrace{\begin{pmatrix}
	-i b_1 -d_1 & - i\alpha a_0\\ i \alpha a_0^* & i b_2 - d_2
	\end{pmatrix}}_{=: L\left(t\right)}
\underbrace{\begin{pmatrix}
	a_1 \\ a_2^*
	\end{pmatrix}}_{=:v\left(t\right)}
\end{equation}
The only possible solution of equation \eqref{eq:stability} in the steady-state \(v_{\mathrm{ss}}\) (where \(\frac{\partial}{\partial t} v_{\mathrm{ss}} = 0\) ) is given by the trivial steady-state solution \(v_{\mathrm{ss}} = 0\) as long as the steady-state stability matrix \(L_{\mathrm{ss}}\) is not invertible. This condition is equivalent to \(\det\left(L\right) = 0\) from which we deduce the equations
\begin{IEEEeqnarray}{rll}
|a_0|^2  & \: = \: & \frac{b_1 b_2 + d_1 d_2}{\alpha^2} \label{eq:stability1}\\
0  & \: = \: & b_1 d_2 - b_2 d_1 \label{eq:stability2}
\end{IEEEeqnarray}
In all the calculations we have used equation \eqref{eq:stability2} to determine the oscillation frequencies \(\omega_{\mathrm{osc},n}\) in terms of the system parameters and modal amplitudes. In order to find the very specific point at which the trivial solution becomes unstable, and the system transitions from the trivial solution to the above threshold solution, it suffices to solve for equations (\ref{eq:stability1},\ref{eq:stability2}) by further setting \(a_1 = a_2 = 0\). In this case, we find for the critical steady-state oscillation amplitude \(a_0^{\mathrm{crit}}\) of the drive mode
%
\begin{equation}
\label{eq:crit_point}
a_0^{\mathrm{crit}} = \frac{\sqrt{d_1 d_2}}{|\alpha|}\sqrt{1 + \frac{\Delta^2}{\left(d_1 + d_2\right)^2}}.
\end{equation}\\
where we have introduced the detuning parameter \(\Delta = \omega_{\mathrm{d}} - \left(\omega_{0,1} + \omega_{0,2}\right)\). In many applications, the actuation of oscillatory MEMS is controlled by a phase-locked loop (PLL) in order to maintain resonant actuation despite any frequency shifts during operation \cite{Nabholz2018, Lee2011}. In this case, it can be shown that 
\(\omega_{\mathrm{d}} = \omega_{0,0} + \beta_{n} | a_0^{\mathrm{crit}} |^2\). 
Thus, we arrive at a formulation for the critical amplitude that is no longer a function of operational parameters such as \(\omega_{\mathrm{d}}\), but only of design parameters\\
\begin{equation}
	a_{0,\mathrm{PLL}}^{\mathrm{crit}}  \! = \! \frac{-2 d_{\mathrm{pb}} \delta\!+\!d_{\mathrm{s}}\left(d_s \alpha^2\!\pm\!\sqrt{d_{\mathrm{s}}^2 \alpha^4\!-\!4 d_{\mathrm{pb}} \left(d_{\mathrm{pb}} + \alpha^2 \delta\right)}\right)}{2 d_{\mathrm{pb}}\beta_0}.
\end{equation}
with the short-hand notations \(d_{\mathrm{pb}} := d_1 d_2 \beta_0\) and \(d_{\mathrm{s}} := d_1 + d_2\) and the design parameter for the linear mode frequencies \(\delta = \omega_{0,0} - \left(\omega_{0,1} + \omega_{0,2}\right)\). 
The analytical expression in equation \eqref{eq:crit_point} yields the critical amplitude and, thus, the critical deflection angle for any micro mirror device as a function of the coupling strength \(\alpha\), the damping factors of the parasitic modes, \(d_1\) and \(d_2\), and in particular the internal resonance frequency condition \(\delta\). These quantities can be thought of as the control handle for the design of micro mirror devices with or without the feature of SPDC. Most importantly, the analytic expression in equation \eqref{eq:crit_point} explains why some of our micro mirror devices, despite being taken from the same design layout, feature SPDC and others display the expected Duffing behaviour, see Fig.\,\ref{fig:comparison_mode_coupling}. Whereas process tolerances in the fabrication of the micro mirrors have only small influence on the damping, the absolute resonance frequencies, and the coupling, they have, in fact, great influence on the internal resonance condition \(\delta\). Thus, the process tolerances also greatly influence the critical drive mode amplitude.    
\\
\begin{figure}[ht]
	\centering
	\includegraphics[width=1.0\linewidth]{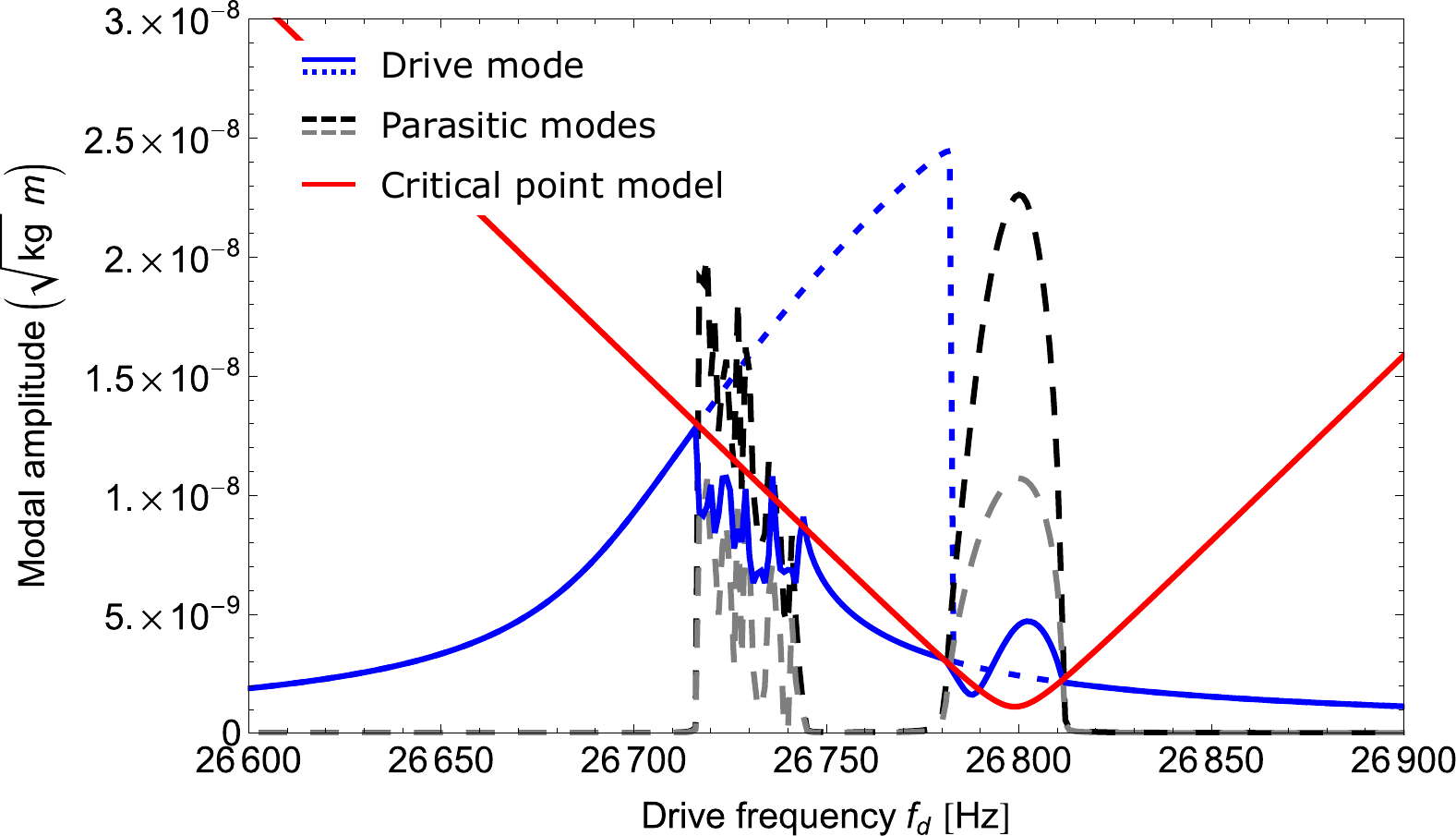}
	\caption{Transiently modelled forward frequency sweeps for all three modes. The red line denotes the critical amplitude, the solution of equation \eqref{eq:crit_point}, for each drive frequency. The solid blue line represents the drive mode for the simulated SPDC coupling, the dotted blue line the drive mode without mode coupling for comparison. The dashed grey and black lines show the behaviour of the two parasitic modes.}
	\label{fig:model_all_modes}
\end{figure}
\subsubsection*{System behaviour above threshold}
Fig.\,\ref{fig:model_all_modes} shows the steady-state amplitudes \(|a_n|\) obtained from the simulated frequency sweep for a parameter set describing the micro mirror device which showed the highest variety of nonlinear effects. In this simulation, we solved the differential equations (\ref{eq:a0}-\ref{eq:a2s}) for a fixed drive frequency up to a time \(T\) much greater than the relaxation time of the system \(\approx \frac{1}{d_0}\). The steady-state value shown in Fig.\,\ref{fig:model_all_modes} was obtained by evaluating the system variables at time \(T\). We performed the sweep by stepping through the drive frequencies \(\omega_{\mathrm{d}}\) in \(1\,Hz\) increments. The initial conditions for each calculation were set to the steady-state values of the previous step in the sweep except for the first step, where the initial condition was set to \(a_n = 0\) for all n. With this algorithm, we emulate the measurement procedure in the experiment. In the transient simulations, we introduced a source term for the parasitic modes. Otherwise, the numerical simulation would not reveal the transition of the trivial solution from stable (below threshold) to unstable (above threshold). This source term was chosen to be finite but small such that the overall steady-state result was not affected by it. \\
\begin{figure}[hb]
	\centering
	{
		\includegraphics[width=1.0\linewidth]{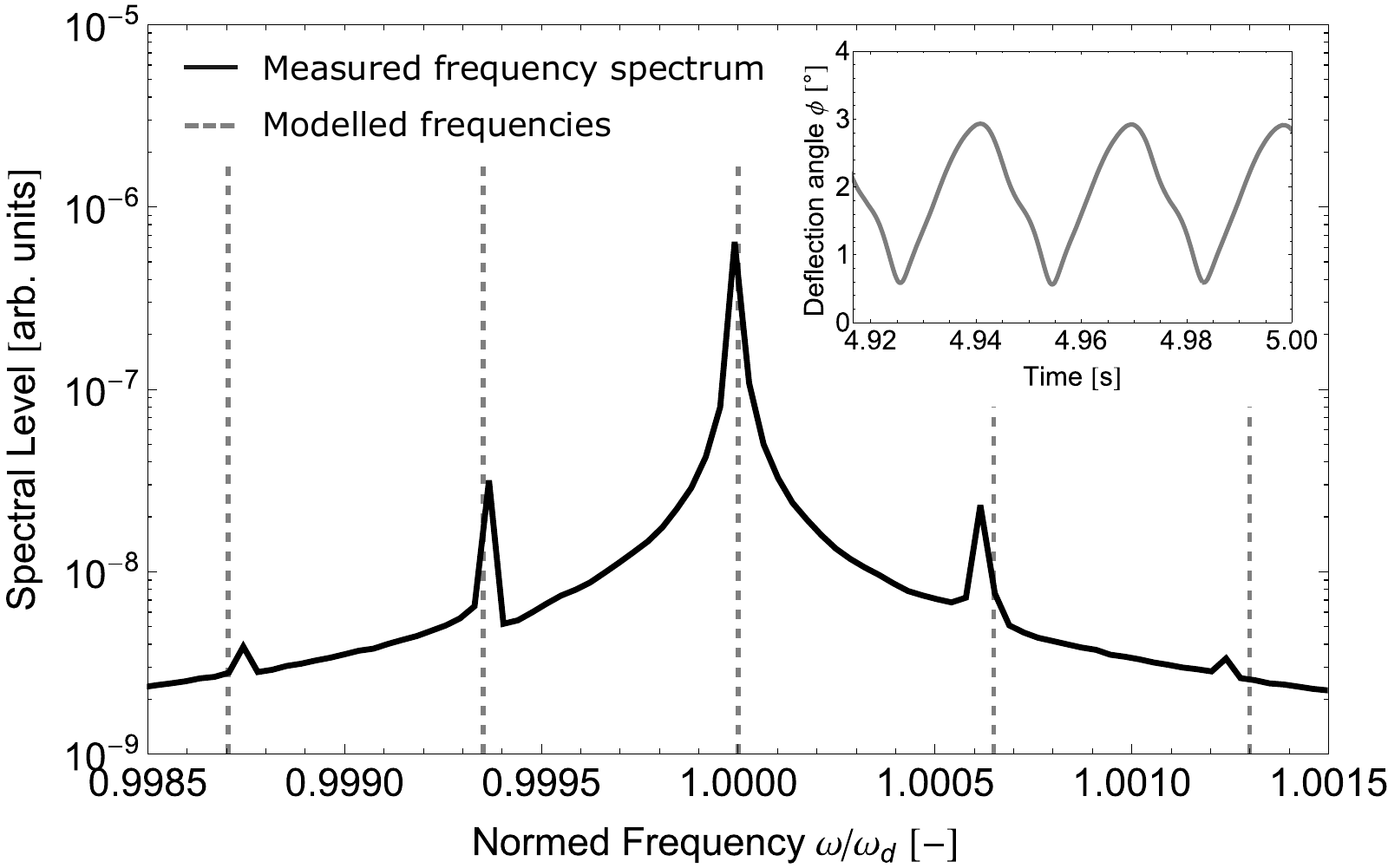}
		\label{SUBFIGURE:freq_combs_comp}
	}
	\caption{Frequency spectrum from measurement of device A contrasted with modelled frequencies during a limit cycle. The measured frequency spectrum is shown in black, whereas the dashed grey lines indicate the modelled frequencies. The signal envelope obtained from RWA (shown in grey in the inset) is analysed and yields a range of slow oscillations, the lowest of which at roughly \(17\,Hz\) can still be observed as flicker by the human eye.}
	\label{fig:fft}	
\end{figure}
\noindent
In Fig.\,\ref{fig:model_all_modes}, the drive mode is depicted by the solid blue line, the parasitic modes 1 and 2 are shown as dashed lines in grey and black, respectively. The red line denotes the critical drive mode amplitudes given by equation \eqref{eq:crit_point}. As a reference, the dotted blue line shows the expected behaviour of the drive mode as if mode coupling did not occur. We note that the critical drive mode amplitude predicted by equation \eqref{eq:crit_point} agrees well with the transient simulation. However, the critical point model at threshold can only predict the limits of the coupling region and does not make a statement about the effects inside the region. \\
Before the drive mode reaches its critical oscillation amplitude, the parasitic modes are not actuated at all. At the transition into the SPDC regime, resonant actuation of the parasitic modes occurs such that the oscillation amplitudes of the parasitic modes compare with the drive mode oscillation amplitude. This actuation is due to an energy transfer from the drive mode to the two parasitic modes. Thus, resonant actuation of the parasitic modes entails amplitude depletion of the drive mode. \\
\noindent
Remarkably, for the specific set of parameters, the system is able to enter and leave the above threshold regime twice during the frequency sweep. The second region (\(f_{\mathrm{d}}\) between \(26780\,Hz\) and \(26812\,Hz\)) above threshold displays stable steady-state solutions, whereas the first region (\(f_{\mathrm{d}}\) between \(26715\,Hz\) and \(26745\,Hz\)) above threshold displays amplitude instabilities in all three modes exhibiting periodic steady-state solutions, i.e. limit cycles or rather amplitude modulations. We have depicted exemplary modulations in the inset in Fig.\,\ref{fig:fft} showing the simulated transient behaviour of the deflection angle. At the time range shown, all transient effects have abated and the signal remains periodic.\\
Moreover, Fig.\,\ref{fig:fft} shows a comparison of the frequency spectrum from measurement and the distinct frequency peaks from the simulation during a limit cycle. The measured (grey) and modelled (black) spectrum are compared close to the drive frequency \(f_{\mathrm{d}}\).\\
\begin{table}[b]
	\begin{tabular}{c|c|c|c}
		\hline 
		\multicolumn{2}{c|}{Mode type} & Drive: \(n\!=\!0\) & Parasitic: \(n\!>\!0\) \\
		\hline \hline
		Parameter & Symbol & \multicolumn{2}{c}{Method}\\
		\hline \hline
		Linear mode & \(f_{0,n}\) & \multicolumn{2}{c}{Direct extraction} \\
		frequency & & \multicolumn{2}{c}{from measurement} \\
		\hline
		Linear quality & \(Q_n\) & \multicolumn{2}{c}{Direct extraction} \\
		factor &  & \multicolumn{2}{c}{ from measurement} \\
		\hline
		Duffing & \(\beta_n\) & Extr. below & Fit above  \\
		coefficient &  &  threshold & threshold \\
		\hline
		Nonlinear & \(Q_{\mathrm{nl},n}\) & Extr. below & Fit above \\
		quality factor &  & threshold & threshold \\
		\hline
		Three-wave  & \(\alpha\) & \multicolumn{2}{c}{Extraction at critical} \\
		coefficient & & \multicolumn{2}{c}{amplitude threshold} \\
		\hline
	\end{tabular}
	\caption{Parameters}
	\label{table}
\end{table}
\noindent
Close to the bifurcation points we have been able to observe another interesting phenomenon in both experiment and simulation known as \textit{critical slowing down} \cite{Degenfeld-Schonburg2015, Degenfeld-Schonburg2016, Strogatz2007}. Critical slowing down refers to the slowing down of the relaxation time of the system upon changing the parameter settings across a bifurcation point such that the final parameter settings lie in the vicinity of the bifurcation point. In our experiment, we have changed the drive frequencies just by a small fraction but such that we crossed the critical point from below to above threshold. Depending on the size of the frequency change (we have tried 0.1 Hz up to 1Hz) we had to wait for several seconds or even minutes for the system to display the characteristic spectrum of SPDC. For comparison, the typical relaxation time is on the order of a few milliseconds. Therefore, we have always been concerned to allow sufficient waiting times in both the simulations and experiments to accurately capture the position of the critical points.
\subsection*{Parameter extraction and model validation}
\label{parameter_extraction}
Table\,\ref{table} shows an overview of all linear and nonlinear parameters needed for the system model. It also indicates the extraction procedure which we have used for each parameter. \\
\noindent
Duffing coefficient and nonlinear quality factor of the drive mode were extracted from measurements below threshold like the ones shown in Fig.\,\ref{fig:comparison_mode_coupling}a.\\
\begin{figure}[h]
	\centering
	\includegraphics[width=1.0\linewidth]{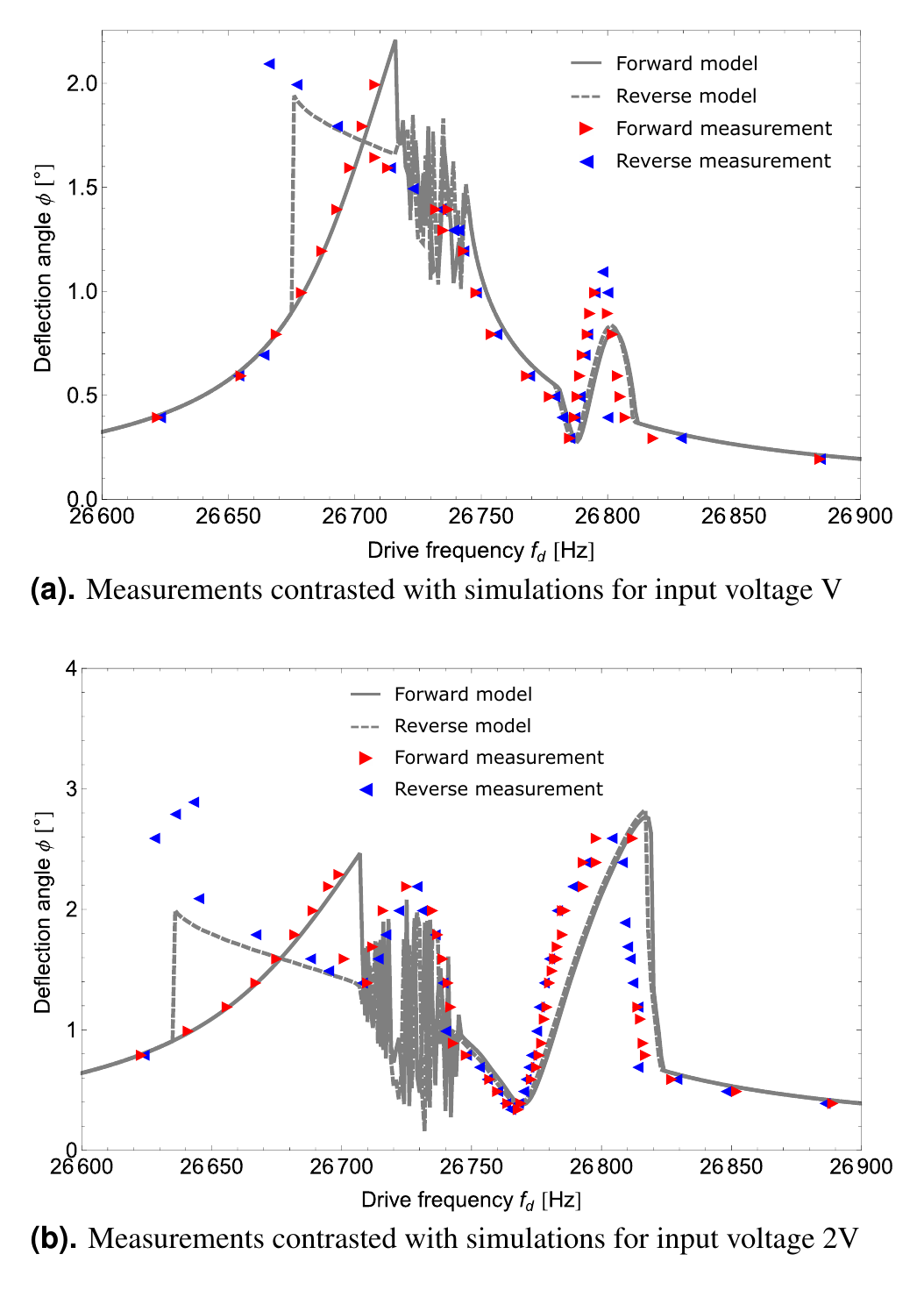}
	\caption{Discrete measurements of device A during frequency sweeps (red triangles denote the forward sweep, blue triangles the reverse sweep) contrasted with the modelled amplitude response curves (solid grey lines for the forward sweep, dashed grey lines for the reverse sweep). Fig.\,\ref{fig:all_in_one}a shows measurements and simulations for an input voltage V, Fig.\,\ref{fig:all_in_one}b for an input voltage 2V. Note that whenever dynamical system behaviour in the form of unstable oscillations or limit cycles occurs in measurements, only the maximum deflection angle for a given frequency is recorded. }
	\label{fig:all_in_one}	
\end{figure} 
Here, it should be noted that these measurements had to be conducted using the functional micro mirror device B that did not show any signs of mode coupling. Consequently, variations in these parameters occur between the devices, but they are assumed to be minor. Yet, this provides a possible explanation for small deviations between the measured and modelled system.\\
\noindent
Given all the linear parameters, as well as the Duffing- and nonlinear damping coefficient of the drive mode, the coupling coefficient \(\alpha\) that determines the coupling strength between the three modes can be extracted using equation \eqref{eq:crit_point} without the need for any fit parameters. \\
From a parameter fit above threshold, the remaining Duffing coefficients and nonlinear quality factors are obtained.\\
\noindent
Fig.\,\ref{fig:all_in_one} shows measured frequency sweeps for the micro mirror device A (red triangles denote the forward sweep, blue triangles the reverse sweep) and the modelled amplitude response curves (solid grey lines for the forward sweep, dashed grey lines for the reverse sweep). Fig.\,\ref{fig:all_in_one}a shows measurements and simulations for an input voltage \(V\), Fig.\,\ref{fig:all_in_one}b for the input voltage \(2 V\). Note that whenever dynamical system behaviour in the form of unstable oscillations or limit cycles occurs in measurements, only the maximum deflection angle for a given frequency is recorded. In Fig.\,\ref{fig:all_in_one}a, measurements for some frequencies have been omitted, since reliable recording of deflection angles was not possible, due to the amplitude modulations in the limit cycle region. (In contrast, for the measurements in Fig.\,\ref{fig:all_in_one}b, the maximum amplitudes of the limit cycle oscillations could be identified unambiguously.) In both figures, it becomes apparent that hysteresis effects occur, showing bistable states.\\
The comparisons displayed in Fig.\,\ref{fig:all_in_one} show that our model predicts the measured system behaviour with high accuracy, even taking into account that the system is quite sensitive to parameter changes. From this we can conclude that, in this specific case, SPDC does not trigger further couplings or lead to the actuation of more parasitic modes but that the observed coupling stays confined to three modes and is thus predictable. \\
We have confirmed the predicted SPDC couplings for micro mirror device A, as well as for several more micro mirror samples not depicted here. Since the distribution of the linear mode frequencies is largely due to variations in the process tolerances, only few regular devices show such coupling effects. Of these, most lie within a similar frequency range leading to similar detuning parameters. In future, using test devices with purposefully chosen process tolerance variations that lead to a range of different detuning parameters, we aim to publish our results for larger samples of micro mirror devices in order to strengthen the validity of our model, especially regarding the prediction of the critical point \cite{Nabholz2019_unpub}. 
We elaborate on the validity of the RWA in the appendix. 
\section*{Discussion}
We have observed SPDC processes in a MEMS micro mirror and attributed them to non-degenerate three-wave mixing induced by geometric nonlinearities. All linear and nonlinear system parameters were either taken directly from LDV measurements, extracted from optical measurements using the trivial solution of our model below threshold, extracted from measurements using the simplified model at threshold or fitted to measurements using the non-trivial solution of our full system model.\\
The accurate results showcase that our assumption of SPDC holds true and that a nonlinear three DOF model explains all the effects observed in measurements of the micro mirror, such as resonant three-mode excitations, amplitude depletion of the drive mode, hysteretic behaviour and limit cycles. Our model not only predicts the frequency range where coupling occurs, but can also be used to emulate the amplitude response curves within the coupling region. \\
Thus, we have established a procedure with low computational cost that predicts the occurrence of mode coupling and its amplitude. Due to the wide variation in linear mode frequencies which stems from the large deviations from nominal process parameters in MEMS surface technologies, such a prediction can point out individual chips with low critical deflection angles as well as provide valuable input for design optimization. Naturally, a highly desirable future goal is to determine the nonlinear coupling coefficients numerically, e.g. by using finite element formulations of the strain energy \cite{Touze2014}, and thus provide simulative predictions of mode coupling during device design.\\
In essence, our work has shown that fundamental nonlinear phenomena such as SPDC occur in mechanical structures where two prerequisites are given: The vibrational modes of the structure must fulfil an internal resonance condition on one hand and have a reasonably large coupling on the other hand. Due to their complex geometry and thus, mode spectrum, our micro mirror designs have a high likelihood of fulfilling these requirements, in contrast to typical design elements such as plates, membranes or beams. Thus, in future work one could aim at designing mechanical structures not only to study further fundamental effects, but also to exploit nonlinearities for innovative actuation and sensing principles. As a further step, one could even create designs in the nanometre scale to reach for quantum mechanical behaviour governed by fundamental nonlinear processes.
\section*{Methods}
\subsection*{Laser Doppler vibrometry measurements} 
By using a piezoelectric vibratory plate below the micro mirror device with a broad-band Gaussian noise spectrum we actuate the linear vibratory modes of the mirror causing deflections in the nanometre range. Depending on the mode shapes, some modes respond better to the vibratory actuation than others. We can thus measure the resulting response spectrum, see Fig.\,\ref{fig:freq_spectrum_ldv}b. The spectrum contains the information about the linear mode frequencies and the linear quality factors of the three relevant modes.\\
In a different test setup, using direct piezoelectric actuation, we can also detect the response spectrum. In this case, we only find one sharp peak exactly at the drive frequency \(f_{\mathrm{d}}\) below the critical deflection angle, whereas above the threshold of SPDC we find three sharp peaks at \(f_{osc,1}, f_{osc,2}\) and \(f_{\mathrm{d}}\) with the exact relation \(f_{osc,1} + f_{osc,2} = f_{\mathrm{d}}\), see Fig.\,\ref{fig:freq_spectrum_ldv}a.\\
The point of measurement, denoted by the red dot in Fig. \ref{fig:mirrorsketch}, is chosen for its location on the frame. This position entails deflections on a similar scale for all three relevant modes which is of an advantage for precise LDV measurements. The choice of points is also limited by the packaging of the mirror that does not expose the full frame structure, but only the central part.   
\subsection*{Optical measurements of amplitude response curves} 
\begin{figure}[htbp]
	\centering
	\includegraphics[width=0.9\linewidth]{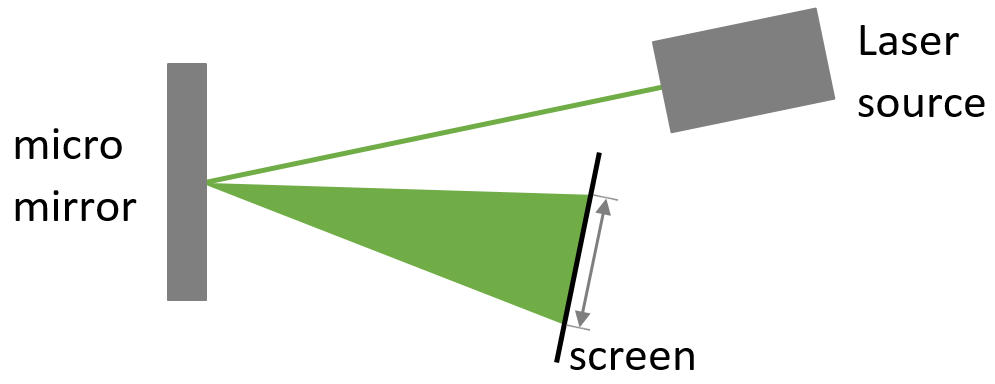}
	\caption{Schematic of the optical test setup: A laser beam that is pointed at the reflective micro mirror surface. When the mirror is actuated, the beam is deflected and projected onto a screen. The length of the resulting line is proportional to the deflection angle of the mirror.}
	\label{fig:opticalsetup}
\end{figure}

In order to conduct frequency sweeps, an optical test setup is used as shown in Fig.\,\ref{fig:opticalsetup}: A laser beam is pointed at the reflective mirror surface and redirected by a stationary mirror onto a screen that incorporates a scale for relating the length of the projected line directly to the deflection angle of the micro mirror (which in turn is directly proportional to the maximum amplitude of the outer mirror edge). 
The actuation of the mirror is provided by a frequency generator that yields a sinusoidal input voltage. The frequency sweep is conducted manually with increments of 1 Hz in forward and reverse direction for each chosen input voltage. Since the onset of parametric down-conversion might experience critical slowing-down, waiting times of several seconds are implemented in the vicinity of the critical point in order to measure stationary state system behaviour. These measurements yield amplitude response curves for forward and reverse frequency sweeps, see Fig.\,\ref{fig:all_in_one}.
\subsection*{Rotating wave approximation}
Rotating wave approximation can be used for the simplified analysis of high-Q oscillatory systems. We introduce the concept of RWA: Starting from equations \eqref{eq:q_n} and \eqref{eq:p_n}, we define an auxiliary function \(f\) as:
\begin{equation}
	\frac{i}{\sqrt{2\omega_{0,n}}} p_n + \sqrt{\frac{\omega_{0,n}}{2}} q_n = a_n e^{-i\omega_{0,n}^{osc}t} =: f.
\end{equation}
The time derivative of \(f\) can be written as
\begin{IEEEeqnarray}{lll}
	\dot{f} & \: = \: & \dot{a}_n e^{-i\omega_{\mathrm{osc},n}t} - i\omega_{\mathrm{osc},n} a_n e^{-i\omega_{\mathrm{osc},n}t} \label{eq:fdot_alternative1}
\end{IEEEeqnarray}
or
\begin{IEEEeqnarray}{lll}
\dot{f} & \: = \: & \frac{i}{\sqrt{2\omega_{0,n}}} \dot{p_n} + \sqrt{\frac{\omega_{0,n}}{2}} \dot{q_n}. \label{eq:fdot_alternative2}
\end{IEEEeqnarray}
We use equations (\ref{eq:3dof_ode0}-\ref{eq:3dof_ode2}) to rewrite \(\dot{p_n} = \ddot{q}_n\): 
\begin{equation}
\dot{p_n} = -\omega_{0,n}^2 q_n - \frac{\omega_{0,n}}{Q_n}\dot{q_n} + F_n \delta_{n,0} \sin\left(\omega_{\mathrm{d}} t\right) - NLT 
\label{eq:pndot}
\end{equation}
with \(\delta_{n,0}\) denoting the Kronecker delta function and the placeholder \(NLT\) denoting the nonlinear terms. We will introduce the averaged terms for the nonlinear terms later. \\
Combining equations \eqref{eq:fdot_alternative1}, \eqref{eq:fdot_alternative2} and \eqref{eq:pndot} and using the identity \(\sin\left(x\right) = \frac{1}{2 i}\left(e^{ix} - e^{-ix}\right)\) yields
\begin{IEEEeqnarray}{lll}
\label{eq:rwa}
	\dot{a}_n e^{-i\omega_{\mathrm{osc},n} t} & \: - \: & i\omega_{\mathrm{osc},n} a_n e^{-i\omega_{\mathrm{osc},n}t} \nonumber \\
	& \: = \: & -i\frac{\omega_{0,n}}{2} \left(a_n e^{-i\omega_{\mathrm{osc},n} t} + a_n^* e^{i\omega_{\mathrm{osc},n}t}\right) \label{eq:rwa_full} \nonumber \\
	& \: - \: & \frac{\omega_{0,n}}{2 Q_n} \left(a_n e^{-i\omega_{\mathrm{osc},n} t} - a_n^* e^{i\omega_{\mathrm{osc},n} t}\right) \nonumber \\ 
	& \: + \: & i \delta_{n,0} \frac{F_n\left(t\right)}{\sqrt{2 \omega_{0,n}}}\cdot \frac{1}{2 i}\left(e^{i\omega_{\mathrm{d}} t}-e^{-i\omega_{\mathrm{d}} t}\right) \nonumber \\
	& \: - \: & i\frac{\omega_{0,n}}{2}\left(a_n e^{-i\omega_{\mathrm{osc},n}t} - a_n^* e^{i\omega_{\mathrm{osc},n}t}\right) - NLT.
\end{IEEEeqnarray}
The RWA is based on the central assumption that fast oscillations can be neglected. In practise, the RWA is performed by multiplying equation \eqref{eq:rwa} on both sides by \(e^{i\omega_{\mathrm{osc},n} t}\) and discarding any fast oscillating terms, i.e. we consider only terms where the exponent of the exponential function equals zero. Keeping in mind that \(\omega_{\mathrm{d}} = \omega_{\mathrm{osc},0} = \omega_{\mathrm{osc},1} + \omega_{\mathrm{osc},2}\), we end up with
\begin{IEEEeqnarray}{lll}
	\dot{a}_n & \: = \: & -i\Delta_n a_n- \frac{\omega_{0,n}}{2 Q_n} a_n - \delta_{n,0} \frac{F_n\left(t\right)}{2\sqrt{2\omega_{0,n}}} - NLT\\
	& \: = \: & -i\Delta_n a_n- \frac{\omega_{0,n}}{2 Q_n} a_n -i P_n^* - NLT \nonumber
\end{IEEEeqnarray}
Here, we introduce \(\Delta_n = \omega_{0,n} - \omega_{\mathrm{osc},n}\).
In the same spirit, we obtain all nonlinear terms in equation \eqref{eq:3dof_ode0}:
\begin{IEEEeqnarray}{lll}
	\beta_n q_n^3 & \: \rightarrow \: & -i \beta_n |a_n|^2 a_n \\
	\frac{\omega_{0,n}}{Q_{\mathrm{nl},n}}q_n^2 \dot{q}_n \ & \: \rightarrow \: & - \frac{1}{4 Q_{\mathrm{nl},n}} a_n |a_n|^2 
\end{IEEEeqnarray}
The three-wave coupling terms take on different forms for the three modes: For the drive mode, we obtain \(\alpha q_1 q_2 \rightarrow -i \alpha a_1 a_2\). Similarly, the parasitic modes 1 and 2 yield \(\alpha q_0 q_2 \rightarrow -i \alpha a_0 a_2^*\) and \(\alpha q_0 q_1 \rightarrow -i \alpha a_0 a_1^*\), respectively.\\
Like in other averaging methods \cite{Strogatz2007}, the validity of the RWA for the three degree-of-freedom model given by equations \eqref{eq:3dof_ode0} - \eqref{eq:3dof_ode2} depends on a time-scale separation between slow and fast variables. Whereas the fast oscillation changes on the time-scale of oscillation frequencies \(\propto \omega^{-1}\), the slow variables change on time-scales \(\propto \left(\frac{\omega}{Q}\right)^{-1}\). The RWA assumes that the fast oscillations average to zero during the transient evolution of the slow variables. Thus, the RWA is valid for high-Q systems, i.e. \(Q \propto 10^2\) or higher. In additon, by neglecting the fast oscillating terms, the RWA implicitly assumes that the remainder of the fast oscillating terms which has not been averaged out is small. For nonlinear terms, this only holds true if the nonlinearities which are the product of nonlinear (coupling) coefficients and amplitude are small in the regimes of interest. One possible way to quantify this statement is by comparing the nonlinear and the linear stiffness: The RWA is valid if e.g. \(\frac{\tilde{\beta}_n |q_n|^2}{\omega_{0,n}^2} \ll 1\) and \(\frac{\tilde{\alpha}|q_n|}{\omega_{0,m}^2} \ll 1\) for all \(n,m \in \{0,1,2\}\). 
\section*{Data availability statement}
The datasets generated during and analysed during the current study are available from the corresponding author on reasonable request.
%

%
%
%
%
\section*{Author contributions statement}
P.D. crafted the original idea, U.N. and P.D. developed the theory, U.N. performed the calculations. F.S. conceived the experiments, U.N. and F.S. conducted the experiments. U.N. wrote the manuscript, P.D. revised the manuscript, all authors reviewed the manuscript. J.M. also provided general guidance. 
\section*{Additional information}
%
\textbf{Competing interests} The authors declare no competing interests. \\
%
%
%
%
%
\end{document}